\documentclass{article}
\usepackage{amssymb}
\usepackage{bm}
\usepackage[dvips]{graphicx}
\begin{document}

\title{Possibility of exchange switching ferromagnet-antiferromagnet junctions}

\author{Yu. V. Gulyaev, P. E. Zilberman\thanks{E-mail: zil@ms.ire.rssi.ru}, E. M. Epshtein
\\ \\
V. A. Kotelnikov Institute of Radio Engineering and Electronics\\
of the Russian Academy of Sciences, Fryazino, 141190, Russia}

\date{}

\maketitle

\abstract{Current flowing is studied in magnetic junctions consisting of a
ferromagnetic metal (FM), antiferromagnetic conductor (AFM) and a
nonmagnetic metal closing the electric circuit. The FM layer with high
anisotropy and pinned spins of the magnetic atoms in the lattice acts as a
spin injector relative to the AFM layer. To obtain resulting magnetization
in the AFM layer, magnetic field is applied, which may be varied to
control the magnetization. The spin-polarized current from the FM layer
creates a torque and makes it possible to switch the magnetization. A
possibility is shown to lower the threshold current density by the orders
of magnitude by means of the magnetic field. }\\
\\

\section{Introduction}\label{section1}
Antiferromagnetic layers are used conventionally in spintronics to
introduce additional magnetic anisotropy and pin the magnetization vector
direction in a magnetic junction. Such an effect is based on the
unidirectional anisotropy phenomenon~\cite{Salanskii}. Recently, interest has been
revived to studying FM-AFM junctions with special attention to the current
effect on the interface processes~\cite{Sankaranarayanan,Wei1,Urazhdin}. Besides, experimental works have
been revealed~\cite{Wei2,Basset} in which the magnetoresistive effect has been observed
in FM-AFM structures with a point contact between the layers under high
current densities (up to $10^9$--$10^{10}$ A/cm$^2$); the effect is due,
apparently, to the \emph{sd} exchange interaction.  The similar effect is
observed in FM junctions with broad applications~\cite{Fert,Grunberg}. Suppositions have
been voiced that such an effect in FM-AFM or AFM-AFM structures will have
interesting features because of absence of demagnetization, in particular.
Finally, opinions have been stated that the \emph{sd} exchange may lead to
the spin transfer from the conduction electrons to the lattice, as in FM
junctions, and cause instability with the magnetization switching. It has
been supposed that nonlinear effects and the current-driven electron spin
polarization may promote such an effect~\cite{Nunez,Swaving,Gomonay1,Gomonay2}.

Thus, the possibility of the current-driven exchange switching FM-AFM
structures remains, apparently, the most incomprehensible in theory and
experiment. The idea of the electron spin transfer to the lattice by the
spin-polarized current has been put forward in famous works~\cite{Slonczewski,Berger}. The
idea turned out to be very fruitful, so that it is interesting to know
whether it is applicable to the FM-AFM structures. An attempt to clear the
matter up is the goal of the present paper.

When we say about the spin transfer ``to the lattice'', it is necessary to
explain which of two (at least!) AFM sublattices spins transfer to. The
basic works~\cite{Slonczewski,Berger}, naturally, do not answer the question. A physical
picture of the spin transfer, according to~\cite{Berger}, consists in precession of
the current polarization vector $\mathbf p$ around the lattice
magnetization vector $\mathbf M$ and gradual decrease in the precession
angle with motion along the current because of the statistical spreading
of the electron velocities. Such a picture looks general enough and must
remain for any model of the magnetic sublattices, including AFM. If such a
picture is applied, then a torque $\mathbf T$ similar to that in~\cite{Slonczewski} is
to be added to the right-hand side of the Landau-Lifshitz (LL) equation
for the AFM magnetization. Note that the torque $\mathbf T$ reveals in LL
equation for the total magnetization $\mathbf M$, rather than the each of
the sublattices $\mathbf M_1,\,\mathbf M_2$.

A formal derivation of the expression for the $\mathbf T$ moment was
proposed by us in Refs.~\cite{Gulyaev1,Epshtein,Gulyaev2}. It should have in mind, however, that
the torque originates in a very thin layer $\lambda_F\sim1$--2 nm thick at
the FM-AFM interface; such a length coincides with the precession damping
lengths of the $\mathbf M$ and $\mathbf p$ vectors~\cite{Slonczewski,Berger}. So $\mathbf T$
moment exists only in that thin layer, not the whole range of the current
interaction with the magnetic lattice. In accordance with Refs.~\cite{Gulyaev1,Epshtein,Gulyaev2},
we take the $\mathbf T$ effect into account by means of the boundary
conditions.

\section{The structure and equations of motion}\label{section2}
\subsection{Conduction electrons in the AFM layer}\label{subsection2.1}
\begin{figure}
\includegraphics{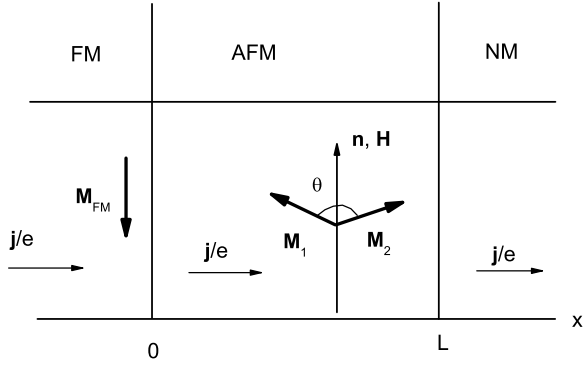}
\caption{The structure in study. The FM layer is a ferromagnetic metal
with $\mathbf{M}_{\rm FM}$ magnetization pinned along $z$ axis; the AFM
layer with $L$ thickness is an easy-plane-type antiferromagnetic conductor with
sublattice magnetization vectors canted by $\theta$ angle in an applied
field $H$ parallel to the anisotropy axis $\mathbf n\|z$;
$|\mathbf{M}_1|=|\mathbf{M}_2|=M_0$. The electric current is closed by a
nonmagnetic conductor NM, so that electron flux $\mathbf{j}/e$ flows
through all the layers.}\label{fig1}
\end{figure}

The structure in study is shown schematically in Fig.~\ref{fig1}. The
sizes along $0y$ and $0z$ axes are large, so that demagnetizaion is absent
in that dimensions. For simplicity, we assume the FM layer to be pinned.
This layer injects spin polarized along $\mathbf
p\equiv\hat{\mathbf{M}}_{\rm FM}$ (here and below the hat over the vector indicates a unit
vector). The main processes occur in the AFM layer. Let us consider the
equations of motion for that layer.

The electron magnetization $\mathbf m(x,\,t)$ obeys the continuity
equation
\begin{equation}\label{1}
  \frac{\partial m_i}{\partial t}+\nabla_kJ_{ik}+g\alpha_{sd}[\mathbf
  m\times\mathbf M]_i-\frac{m_i-\bar m_i}{\tau}=0,
\end{equation}
where $g$ is the gyromagnetic ratio, $\alpha_{sd}$ is the \emph{sd}
exchange constant,
\begin{equation}\label{2}
  J_{ik}\equiv\hat M_i\frac{\mu_B}{e}(j_k^+-j_k^-)=\hat
  M_i\frac{\mu_B}{e}\left\{Pj_k-enD\frac{\partial P}{\partial x_k}\right\}
\end{equation}
is the electron spin current, $j_k^\pm$ are the partial electric current
densities in the spin subbands, $j_k=j_k^++j_k^-$ is the total current
density, $D$ is the spin diffusion constant (it is assumed to be the same
in both subbands), $m_i=m\hat M_i$, $m=\mu_BnP$, $n$ is the electron
density, $P$ is the spin polarization, $\bar m_i=\mu_Bn\bar P\hat M_i$,
\begin{equation}\label{3}
  \bar
  P=\frac{\epsilon_F^{3/2}-\left(\epsilon_F-\epsilon_{sd}\right)^{3/2}}{\epsilon_F^{3/2}+
  \left(\epsilon_F-\epsilon_{sd}\right)^{3/2}}
\end{equation}
is the equilibrium spin polarization, $\epsilon_{sd}$ is the spin subband
energy shift, $\epsilon_F$ is the Fermi energy relative to the bottom of
the lower subband.

The ways of the Eq.~(\ref{1}) transformation and simplification were
discussed in detail in Refs.~\cite{Gulyaev1,Epshtein,Gulyaev2}, so that we present only some
necessary results here.

We consider only slow enough processes due to the vector $\mathbf M$
precession with $\omega\ll 1/\tau$ frequencies. A typical spin relaxation
time in AFM is $\tau\sim10^{-12}$ s. Therefore, the electrons have time to
follow the precession, so that the time derivative may be omitted. In
derivation of Eq.~(\ref{2}), the current density may be assumed to be low,
namely, $j/j_D\ll1$, where $j_D=enl/\tau\sim2\times10^{10}$ A/cm$^2$. This
condition is assumed to be fulfilled. As a result, Eq.~(\ref{1}) is reduced
(see~\cite{Gulyaev1,Epshtein,Gulyaev2}) to
\begin{equation}\label{4}
  \frac{\partial^2m}{\partial x^2}+\frac{m-\bar m}{l^2}=0.
\end{equation}
At the layer boundaries $x=0$ and $x=L$ the solution must obey the
continuity conditions for the spin currents and the differences of the
Fermi quasilevels $(\epsilon_F^+-\epsilon_F^-)$. Then the solution of
Eq.~(\ref{4}) takes the form (see~\cite{Gulyaev2})
\begin{equation}\label{5}
  \Delta m\equiv m-\bar
  m=\mu_BnQ_{\rm FM}\frac{j}{j_D}\frac{\nu\cos\chi}{\nu^\ast+\cos^2\chi},
\end{equation}
where $Q_{\rm FM}=(j_{\rm FM}^+-j_{\rm FM}^-)/j$ is the current polarization in the FM
layer, $\cos\chi=\hat{\mathbf{M}}_{\rm FM}\cdot\hat{\mathbf{M}}(0)$,
$\nu=Z_{\rm FM}/Z_{\rm AFM}$, $\nu^\ast=Z_{\rm FM}/Z_{\rm NM}+\lambda Z_{\rm FM}/Z_{\rm AFM}$,
$\lambda=L/l\ll1$, $Z_q=\displaystyle\frac{l_q\rho_q}{1-Q_q^2}$ are the spin
resistances of the layers, $l_q,\,\rho_q,\, Q_q$ being the spin diffusion
length, the resistivity and the current polarization, respectively, in the
$q=\rm{FM,\,AFM,\,NM}$ layers. Equation~(\ref{5}) describes distribution
of the electron magnetization across the AFM layer.

\subsection{Magnetic sublattices in the AFM layer}\label{subsection2.2}
We derive the equations of motion for the $\mathbf M_1$ and $\mathbf M_2$ sublattice
magnetizations by means of the procedure described in Ref.~\cite{Akhiezer}. The
desired equations take the following general form:

\begin{equation}\label{6}
  \frac{\partial\mathbf M_i}{\partial t}=g[\mathbf
  M_i\times\tilde{\mathbf{H}}_i]+\mathbf R_i\quad(i=1,\,2),
\end{equation}
where the effective fields are determined with variational derivatives of
the AFM energy $W$:
\begin{equation}\label{7}
  \tilde{\mathbf{H}}_i(x,\,t)=-\frac{\partial
  W}{\partial\mathbf{M}_i(x,\,t)}\quad(i=1,\,2);
\end{equation}
the energy is $W=\int w(\mathbf M_1,\,\mathbf M_2)\,d^3x$, the energy
density is $w=w_{ex}+w_{sd}+w_a+w_H$,
$$w_{ex}=\frac{1}{4}\delta(M^2-L^2)+\frac{1}{4}(\alpha+\alpha')
\left(\frac{\partial\mathbf M}{\partial x}\right)^2
+\frac{1}{4}(\alpha-\alpha')
\left(\frac{\partial\mathbf L}{\partial x}\right)^2$$
is the sublattice exchange interaction energy density,
$w_{sd}=-\alpha_{sd}\mathbf m\cdot\mathbf M$ is the \emph{sd} exchange
interaction energy density,
$$w_a=-\frac{\beta}{4}\left((\mathbf n\cdot\mathbf M)^2+(\mathbf n\cdot\mathbf
L)^2\right)-\frac{\beta'}{4}\left((\mathbf n\cdot\mathbf M)^2-(\mathbf n\cdot\mathbf
L)^2\right)$$
is the anisotropy energy density (the anisotropy axis $\mathbf n$ is shown
in Fig.~\ref{fig1}), $w_H=-\mathbf H\cdot\mathbf M$ is the Zeeman energy
density in the external field.

The relaxation terms may be taken in the Gilbert form~\cite{Gomonay1} $$\mathbf
R_i=\frac{\alpha_G}{M_0}\left[\mathbf M_i\times
\frac{\partial\mathbf M_i}{\partial t}\right]\quad(i=1,\,2),$$
$|\mathbf M_1|=|\mathbf M_2|=M_0$.

It is convenient to introduce total magnetization $\mathbf M=\mathbf M_1+\mathbf
M_2$ and antiferromagnetism vector $\mathbf L=\mathbf M_1-\mathbf
M_2$ instead of sublattice vectors $\mathbf M_1,\,\mathbf M_2$. We assume
sublattices to be equivalent. Therefore, a nonzero magnetization $\mathbf
M\ne0$ appears only under applied field $\mathbf H$ parallel to $\mathbf n$ axis,
as shown in Fig.~\ref{fig1}. The $\theta$ angle between the sublattices
and $\mathbf n$ axis is determined by $\mathbf H$ field with
$\cos\theta=H/H_{ex}$, $H\le H_{ex}=2\delta M_0$. The $\theta$ angle is
related with the squares of the vector lengths:
\begin{equation}\label{8}
  M^2-L^2=4M_0^2\cos\theta,\quad M^2+L^2=4M_0^2.
\end{equation}
It is seen that the angle is conserved with motion if the vector lengths
conserve. The magnetization is assumed to be low enough, $M\equiv|\mathbf M|\ll
M_0$. For simplicity, we neglect demagnetization and weak relativistic
noncollinear Dzyaloshinskii interaction~\cite{Akhiezer}.

The equations for $\mathbf M$ and $\mathbf L$ vectors take the form
\begin{equation}\label{9}
  \frac{\partial\mathbf M}{\partial t}=g\left[\mathbf
  M\times\mathbf{H}_{\rm eff}^{\rm FM}\right]+g\left[\mathbf
  L\times\mathbf{H}_{\rm eff}^{\rm AFM}\right]+\mathbf R_1+\mathbf R_2,
\end{equation}
\begin{equation}\label{10}
  \frac{\partial\mathbf L}{\partial t}=g\delta\left[\mathbf
  M\times\mathbf L\right]+g\left[\mathbf
  M\times\mathbf{H}_{\rm eff}^{\rm AFM}\right]+g\left[\mathbf
  L\times\mathbf{H}_{\rm eff}^{\rm FM}\right]+\mathbf R_1-\mathbf R_2.
\end{equation}
New effective fields appear here:
\begin{equation}\label{11}
    \mathbf{H}_{\rm eff}^{\rm FM}=\frac{1}{2}(\alpha+\alpha')\frac{\partial^2\mathbf
  M}{\partial x^2}+\mathbf H+\mathbf{H}_a^{\rm FM}+\mathbf{H}_{sd},
\end{equation}
\begin{equation}\label{12}
    \mathbf{H}_{\rm eff}^{\rm AFM}=-\frac{1}{2}(\alpha-\alpha')\frac{\partial^2\mathbf
  L}{\partial x^2}+\mathbf{H}_a^{\rm AFM}
\end{equation}
with anisotropy fields
$$\mathbf{H}_a^{\rm FM}=\frac{1}{2}(\beta+\beta')(\mathbf n\cdot\mathbf
M)\mathbf n,\quad\mathbf{H}_a^{\rm AFM}=\frac{1}{2}(\beta-\beta')(\mathbf n\cdot\mathbf
L)\mathbf n.$$
The relaxation terms contain vector products $\left[\mathbf M\times\partial\mathbf
M/\partial t\right]$, $\left[\mathbf L\times\partial\mathbf
L/\partial t\right]$, and $\left[\mathbf M\times\partial\mathbf
L/\partial t\right]$, $\left[\mathbf L\times\partial\mathbf
M/\partial t\right]$. By transforming these products by means of
Eqs.~(\ref{9}) and~(\ref{10}) with Eqs.~(\ref{11}) and~(\ref{12}) taking
into account, we find that the products are of the same order of
magnitude, so that we may put $\mathbf{R}_1\pm\mathbf{R}_2\equiv
\mathbf{R}=\displaystyle\frac{2\alpha'_G}{M_0}\left[\mathbf{M}
\times\displaystyle\frac{\partial\mathbf{M}}{\partial
t}\right]$ and use the Gilbert formula with $\alpha'_G\sim\alpha_G$. The
\emph{sd} exchange effective field is
\begin{equation}\label{13}
  \mathbf{H}_{sd}(x,\,t)=-\frac{\partial
  w_{sd}}{\partial\mathbf{M}(x,\,t)}=H_{eq}(M)\hat{\mathbf{M}}+\Delta\mathbf{H}_{sd}.
\end{equation}
The first term in the right-hand side of Eq.~(\ref{13}) is the equilibrium
contribution that directed along $\mathbf{M}$ vector and vanishes from
Eq.~(\ref{9}). The second term is due to the current. It determines with
the nonequilibrium part $\Delta m$ of the electron magnetization. This
term coincides with the one calculated in Refs.~\cite{Epshtein,Gulyaev2}. So we may use
those results here. We have
\begin{equation}\label{14}
  \Delta\mathbf{H}_{sd}=\hat{\mathbf{M}}_{\rm FM}h_{sd}l\delta(x-0),
\end{equation}
\begin{equation}\label{15}
  h_{sd}=(\alpha+\alpha')\mu_BnQ_{\rm FM}\lambda\frac{j}{j_D}\frac{\nu(\nu^\ast-\cos^2\chi)}
  {(\nu^\ast+\cos^2\chi)^2}.
\end{equation}

According to Eqs.~(\ref{11}) and~(\ref{12}), the processes in the AFM
layer differ substantially from those in the FM layer. In particular, even
the lengths of the $\mathbf{M}$ and $\mathbf{L}$ vectors do not conserve
in motion. The possibility of the electron spin transfer to the lattice
similar to that in Refs.~\cite{Slonczewski,Berger} is to be discussed anew. At the same
time, the electron interaction with the lattice is described with
$\mathbf{H}_{sd}$ field and Eqs.~(\ref{14})--(\ref{12}), as in the FM
layer. Therefore, it is necessary only, that two magnetic sublattices
precess in synchronism as in ferromagnets.

Such a synchronism is attained easier if the sublattices are the same. Let
the following conditions are fulfilled:
$\alpha+\alpha'\gg|\alpha-\alpha'|\to0$ and
$\beta+\beta'\gg|\beta-\beta'|\to0$. The relaxation constant is put to be
small, too ($\alpha_G\to0$). The uniform exchange dominates in the
equations because of the following estimates~\cite{Akhiezer}:
$\delta\sim\alpha/a^2\sim10^4\gg1$, $\alpha\sim10^{-12}$ cm$^2$,
$a\sim10^{-8}$ cm. Besides, we have typical estimates
$\alpha_{sd}\sim10^4$, $M\sim10^2$ G. As a result, Eqs.~(\ref{9})
and~(\ref{10}) are reduced to
\begin{equation}\label{16}
  \frac{\partial\mathbf M}{\partial t}=g\left[\mathbf
  M\times\mathbf{H}_{\rm eff}^{\rm FM}\right]+\mathbf R,\quad
  \frac{\partial\mathbf L}{\partial t}=g\delta\left[\mathbf
  M\times\mathbf{L}\right].
\end{equation}

These equations describe precession of the magnetization vector and the
precession-driven forced oscillation of the antiferromagnetism vector.
Note that the vector squares $M^2$ and $L^2$ conserve in accordance
with~(\ref{16}). So it follows from~(\ref{8}) that the sublattice canting
angle $\theta$ conserves also in motion that is substantial in our model.

\section{Spin currents and boundary conditions}\label{section3}
Let us derive expressions for the spin currents in the lattice. According
to Refs.~\cite{Gulyaev1,Epshtein,Gulyaev2}, these expressions follow from the equations of motion.
However, the second of Eqs.~(\ref{16}) does not contain nonlocal or
singular terms. Therefore, the currents may be derived from the first
equation only. We obtain from~(\ref{16})
\begin{equation}\label{17}
  g\alpha\left[\mathbf M\times\frac{\partial^2\mathbf M}{\partial
  x^2}\right]=a\frac{\partial}{\partial x}\left[\hat{\mathbf{M}}\times
  \frac{\partial\mathbf M}{\partial x}\right]\equiv\frac{\partial\mathbf{J}_M}{\partial x}
\end{equation}
with a magnetization flux
\begin{equation}\label{18}
  \mathbf{J}_M=a\left[\hat{\mathbf{M}}\times
  \frac{\partial\mathbf M}{\partial x}\right],
\end{equation}
where $a=g\alpha M$ has the meaning of a magnetization diffusion constant.

Let us consider the singular term due to \emph{sd} exchange in the first
of Eqs.~(\ref{16}), namely, $g\left[\mathbf M\times\mathbf{H}_{sd}\right]$.
By substituting~(\ref{14}) and~(\ref{15}) to it, we have
\begin{equation}\label{19}
  g\left[\mathbf M\times\mathbf{H}_{sd}\right]=\frac{\partial\mathbf{J}_{sd}}{\partial
  x},
\end{equation}
where a \emph{sd} exchange flux appears
\begin{equation}\label{20}
  \mathbf{J}_{sd}=gh_{sd}l\left[\mathbf
  M(0)\times\hat{\mathbf{M}}_{\rm FM}\right]\theta(x-0),
\end{equation}
$\theta(x)$ being the Heaviside step function.

Now let us return to the electron spin current $J_{ik}$ (see
Eq.~(\ref{2})). Here $k=x$, and the current exists in two magnetic layers,
FM and AFM. However, the spatial gradient of the spin polarization $P$ is
absent in the FM layer, so that only the first term proportional to the current density $j$
remains in Eq.~(\ref{2}). The spin current itself contains spins collinear
to the FM layer magnetization, i.e.,
\begin{equation}\label{21}
  \mathbf{J}^{\rm FM}\equiv\mathbf{J}(-0)=\hat{\mathbf{M}}^{\rm FM}\frac{\mu_B}{e}P^{\rm FM}j.
\end{equation}

The spin current continuity conditions at $x=0$ and $x=L$ interfaces give
the boundary conditions necessary to solve Eqs.~(\ref{16}). As the lattice
is pinned in the FM layer, we have $\mathbf{J}^{\rm FM}_M\equiv\mathbf{J}_M(-0)=0$
and $\mathbf{J}^{\rm FM}_{sd}\equiv\mathbf{J}_{sd}(-0)=0$. Then we obtain for
$x=0$
\begin{equation}\label{22}
    \mathbf{J}(+0)-\mathbf{J}(-0)+\mathbf{J}_M(+0)+\mathbf{J}_{sd}(+0)=0.
\end{equation}
Let us project Eq.~(\ref{22}) to the $\hat{\mathbf{M}}$ direction and to
plane perpendicular to that direction. Two equations are obtained:
\begin{eqnarray}\label{23}
    &&\hat{\mathbf{M}}\left(\hat{\mathbf{M}}\cdot\mathbf{J}(-0)\right)=\mathbf{J}(+0),\nonumber
    \\ &&\left[\hat{\mathbf{M}}\times\left[\mathbf{J}(-0)\times\hat{\mathbf{M}}\right]\right]=
    \mathbf{J}_M(+0)+\mathbf{J}_{sd}(+0),
\end{eqnarray}
which are the desired boundary conditions. At $x=L$ boundary, the
continuity condition gives instead of~(\ref{22})
\begin{equation}\label{24}
    \mathbf{J}(L+0)-\mathbf{J}(L-0)-\mathbf{J}_M(L-0)=0.
\end{equation}
After projecting to $\hat{\mathbf{M}}$, two conditions are obtained:
\begin{equation}\label{25}
    \mathbf{J}(L+0)=\mathbf{J}(L-0),\quad\mathbf{J}_M(L-0)=0.
\end{equation}

To express~(\ref{23}) and~(\ref{25}) conditions explicitly as requirements
imposed on the desired magnetization, we multiply the conditions by
$\hat{\mathbf{M}}$ vectorially. Then we obtain finally
\begin{equation}\label{26}
  \frac{\partial\hat{\mathbf{M}}(x)}{\partial x}\Bigl|_{x=0}=
  k\left[\hat{\mathbf{M}}^{\rm FM}\times\hat{\mathbf{M}}(0)\right],
\end{equation}
\begin{equation}\label{27}
  \frac{\partial\hat{\mathbf{M}}(x)}{\partial x}\Bigl|_{x=L}=0
\end{equation}
with a parameter
\begin{equation}\label{28}
  k=\frac{\mu_BQ_{\rm FM}}{aM}\frac{j}{e}\frac{\nu^\ast}{\nu^\ast+\cos^2\chi},
\end{equation}
which characterizes the torque value.

\section{The macrospin approximation}\label{section4}
The problem is simplified considerably if the AFM layer is thin enough, so
that $\lambda\ll1$, $L\ll\sqrt{\alpha/\beta}$ conditions are fulfilled and
the magnetization is almost constant in that layer. Then the following
expansion is valid:
\begin{equation}\label{29}
    \hat{\mathbf{M}}(x)=\hat{\mathbf{M}}(0)+\hat{\mathbf{M}}'(0)x
    +\frac{1}{2}\hat{\mathbf{M}}''(0)x^2+\ldots,
\end{equation}
where the primes mean derivatives with respect to $x$ coordinate. Thus,
several functions of a single variable (time $t$) are introduced instead
of a single $\hat{\mathbf{M}}(x,\,t)$ function of two variables. It means
physically, that a single large domain (``macrospin'') is placed in the
AFM layer~\cite{Slonczewski}. Such an approximation often corresponds to experiments and
simplifies calculations.

By differentiating~(\ref{29}) with respect to $x$, substituting $x=L$ and
using boundary conditions~(\ref{26}) and~(\ref{27}), we obtain
\begin{equation}\label{30}
    \hat{\mathbf{M}}''(0)=-L^{-1}\hat{\mathbf{M}}'(0)=-\frac{k}{L}\left[\hat{\mathbf{M}}^{\rm FM}
    \times\hat{\mathbf{M}}\right].
\end{equation}
The only term in Eq.~(\ref{16}) that contains the second derivative may be
transformed with~(\ref{30}) taking into account:
\begin{equation}\label{31}
    a[\hat{\mathbf{M}}(0)\times\hat{\mathbf{M}}''(0)]=k[\hat{\mathbf{M}}(0)
    \times[\hat{\mathbf{M}}(0)\times\hat{\mathbf{M}}^{\rm FM}]].
\end{equation}
By substituting~(\ref{31}) into Eq.~(\ref{16}), we have
\begin{eqnarray}\label{32}
  &&\frac{d\hat{\mathbf{M}}(0)}{dt}+g[\hat{\mathbf{M}}(0)\times\mathbf H']+
  \frac{ak}{L}[\hat{\mathbf{M}}(0)
    \times[\hat{\mathbf{M}}(0)\times\hat{\mathbf{M}}^{\rm FM}]]-\nonumber \\
    &&-\alpha'_G\left[\hat{\mathbf{M}}(0)\times\frac{d\hat{\mathbf{M}}(0)}{dt}\right]=0,
\end{eqnarray}
where $\mathbf H'=\mathbf H+\mathbf{H}_a^{\rm FM}$.

At first sight, Eq.~(\ref{32}) coincides practically with that in~\cite{Slonczewski} that
was reproduced in many publications. There is an important difference,
however. In Eq.~(\ref{2}) \emph{sd} exchange interaction is absent because
of fulfilling condition~(\ref{14}), so that the equation is valid, in principle,
at any distance from $x=0$ plane where torque $\mathbf T$ originates. At
$x>\lambda_F$, vectors $\mathbf{M}_{\rm FM}$ and $\mathbf{M}$ become
collinear, so the torque vanishes. However, the spin current created in the lattice by
the torque remains.

\section{Exchange instability conditions}\label{section5}
Let us discuss solution of the linearized dynamical equation~(\ref{32})
for small harmonic fluctuations
$\Delta\hat{M}_x,\,\Delta\hat{M}_y\sim\exp(-i\omega t)$ near the
stationary state $\bar{\hat{\mathbf{M}}}=\hat{\mathbf{z}}$. The dispersion
equation takes the form
\begin{equation}\label{33}
  \omega^2+2i\nu\omega-w=0,
\end{equation}
where the following notations are used:
\begin{equation}\label{34}
  w=\frac{1}{1+\alpha'^2_G}\left[\Omega_x\Omega_y+\left(\frac{ak}{L}\right)^2\right],
\end{equation}
\begin{equation}\label{35}
  \nu=\frac{\alpha'_G}{1+\alpha'^2_G}\left[\frac{1}{2}(\Omega_x+\Omega_y)+
  \frac{ak}{L\alpha'_G}\left(\bar{\hat{\mathbf{M}}}\cdot\mathbf
  n\right)\right],
\end{equation}
$\Omega_x=g((\mathbf{H}\cdot\hat{\mathbf{M}})+H_a^{\rm FM}+4\pi M)$,
$\Omega_y=g((\mathbf{H}\cdot\hat{\mathbf{M}})+H_a^{\rm FM})$.

The general instability condition $\rm{Im}\omega>0$ requires fulfilling
inequality $\nu<0$. Put, for example, the following typical parameter
values: $\Omega_x\approx4\pi gM\sim10^{11}$ c$^{-1}$,
$\Omega_y=g(H+H_a^{\rm AFM})\sim10^9$ c$^{-1}$,
$\alpha'_G\approx10^{-1}\ll1$. Then $\nu<0$ condition is approximately
reduces to the following inequality:
\begin{equation}\label{36}
  1+\frac{2ak}{\alpha'_GL\Omega_x}\left(\bar{\hat{\mathbf{M}}}\cdot\mathbf
  n\right)<0.
\end{equation}

By solving the inequality~(\ref{36}) with respect to the current density
and taking the definition~(\ref{28}) into account, we obtain the following
instability condition:
\begin{equation}\label{37}
  -\left(\bar{\hat{\mathbf{M}}}\cdot\mathbf
  n\right)\frac{j}{e}>\lambda\alpha'_Gl\frac{2\pi gM^2}{\mu_BQ_{\rm
  FM}}\left(1+\frac{1}{\nu^\ast}\right).
\end{equation}
Note that we discuss only one current direction when the electrons flow
from the FM layer towards to the AFM layer. It is seen from~(\ref{37})
that the instability occurs only at $\left(\bar{\hat{\mathbf{M}}}\cdot\mathbf
  n\right)<0$ under such a current direction. As $\mathbf n$ vector is
  co-directed with $\bar{\hat{\mathbf{M}}}_{\rm FM}$, it is clear that two
  magnetizations $\bar{\hat{\mathbf{M}}}_{\rm FM}$ and
  $\bar{\hat{\mathbf{M}}}$ should be antiparallel to create instability.

  It follows from~(\ref{37}) that the threshold current density is
  proportional to the dissipation constant and squared magnetization. The
  latter is determined with the field and can be decreased. For example,
  if $M\sim10$ G is taken, then the threshold current density decreases by
  two orders of magnitude in comparison with the typical value for the FM
  junctions.

\section{Conclusions}\label{section6}
The present work has preliminary character, however, it allows to make some
conclusions:
\begin{itemize}
  \item FM--AFM structure may be canted in a magnetic field, i.e., a
  resulting magnetization can be induced. It allows to consider an exchange
  switching by means of a current-driven torque similar to that in a FM
  junction.
  \item To decrease the threshold current, the AFM layer must have minimal
  dissipation and minimal value of the field-induced magnetization.
  \item The instability is possible under antiparallel relative
  orientation of the FM and AFM layers.
\end{itemize}

\section*{Acknowledgments}
The authors are grateful to Prof. G.~M.~Mikhailov for useful discussions.

The work was supported by the Russian Foundation for Basic Research,
Grant No.~10-02-00030-a.

\end{document}